\documentclass{article}
\usepackage{changepage} 
\usepackage[a4paper, left=1.5cm, right=1.5cm, top=2.5cm, bottom=2.5cm]{geometry}
\usepackage{float}
\usepackage{lipsum}
\usepackage{xcolor}
\usepackage{graphicx}
\usepackage[round]{natbib}
\usepackage{pgfplots}
\usepackage{algorithm}
\usepackage{algorithmic}
\usepackage{pgfplotstable}
\pgfplotsset{compat=1.18}
\usepackage{breqn}
\usepackage{xurl}
\usepackage{mathtools}

\usepackage[colorlinks=true,
            linkcolor=blue,
            citecolor=blue,
            urlcolor=blue]{hyperref}

\usepackage[normalem]{ulem}

\usepackage[nameinlink]{cleveref}

\begin{document}

\hypersetup{
    colorlinks=true,
    linkcolor=blue,
    citecolor=blue,
    urlcolor=blue,
    pdfborderstyle={/S/U/W 1} 
}
\twocolumn[
  \begin{@twocolumnfalse}
    \begin{adjustwidth}{-.3cm}{-.3cm}
      \begin{center}
        \vspace{8em}

        \huge \textbf{Aster: A Flexible and Robust Collection of Numerical Tools for N-Body Problems}

        \vspace{2em}

        \large Federico Benevelli

        \vspace{1em}

        \small September 6, 2026

        \vspace{2em}
      \end{center}

    \end{adjustwidth}
    \begin{abstract}
The Aster library is an open-source collection of numerical tools designed for large-scale gravitational N-body simulations. Aster provides a flexible environment to its users, enabling it to be used in the analysis of a broad range of astrophysical problems. The library implements efficient force calculation algorithms, including a spatial tree solver optimized with coordinate dependent identifiers, which has been fine-tuned for both CPU and GPU architectures. Despite its high degree of modularity, many parts of the library have been optimized with \texttt{SIMD} instructions, heavy parallelization, and support on dedicated hardware. Aster also offers a wide variety of numerical symplectic integrators, such as SABA (from first to tenth order), the classic leapfrog algorithm, and the more advanced Wisdom-Holman (WH) integrator, while also providing a vast collection of mathematical tools, e.g., post-newtonian expansions (1PN, 2PN, 2.5PN), data collection and graphing systems, a built-in renderer,  and presets for faster instantiation of large amounts of bodies. In this work we will discuss in depth the details of each module as well as the results of several performance- and accuracy-related tests, probing the library's capabilities in different physical applications.
    \end{abstract}
  \end{@twocolumnfalse}
]
\section{Introduction}
Despite the recent advances in algorithms, the long-term evolution of many-body systems remains a computationally demanding problem by nature. From the evolution of planetary rings comprised of millions of particles to the dynamics of stellar clusters and galaxy formation, modern astrophysical research demands the ability to simulate complex many-body systems over large timescales while maintaining a sufficient numerical accuracy for physically meaningful results. Aster is built as a suite of numerical tools designed to address these issues and optimized for efficient execution on CPU and dedicated hardware, resulting in an ecosystem that is both capable of offering great flexibility over a broad range of physical systems and also of maintaining high numerical accuracy without sacrificing performance.

The library provides two force evaluation routines: a direct $\mathcal{O}(N^2)$ solver, useful when near machine precision is needed in few-body and moderate-scale systems; and an approximate Barnes-Hut algorithm \citep{Barnes1986} with O(N log N) complexity. Both solvers have been implemented with extensive parallelization using \texttt{TBB}, \texttt{SIMD} vectorization, and GPU acceleration via \texttt{OpenCL}.

To ensure numerical stability over long timescales and high particle counts, Aster implements a suite of integration methods designed to preserve the system’s symplectic structure. These include the SABA integrator (up to tenth order), the Wisdom-Holman method for hierarchical systems, and lower-order schemes such as leapfrog and symplectic Euler. Aster also provides post-Newtonian terms (1PN, 2PN, 2.5PN) to account for high-energy regimes where General Relativity's effects are not negligible, as well as an automated data collection system, a real-time 3D renderer equipped with an API capable of displaying custom data, and utilities for efficient instantiation of large-scale structures. In the following sections, we describe Aster's design choices and numerical methods in detail. We present performance benchmarks demonstrating the library's capabilities across various regimes.
    
\section{Code Overview}
The Aster library is entirely written in \texttt{C++} and is available under an open-source license. Its source code can be found at \citet{Github} and installed using the source control software \texttt{git}. The code compiles on modern Unix-like systems, such as Linux and macOS, and requires the libraries \texttt{OpenGL}, \texttt{OpenCL}, and \texttt{TBB} to be compiled and executed correctly.

\subsection{Submodules} Aster comes equipped with various numerical integrators, such as SABA from \citet{Laskar2001}, Wisdom-Holman (WH) \citep{Rein2012}, leapfrog, and Euler, as well as force evaluation or approximation routines implemented for CPU and dedicated hardware. Additionally, a 3D renderer is included in the installation, which can show a simulation's content in various levels of detail  and allows users to show data in real time using a built-in API. Moreover, an automatic data-collection API, which is capable of saving to disk arbitrary data (e.g., position of the bodies over time, total energy, total numerical inaccuracy), is also provided to the users.

\subsection{Installation}
A snapshot of the latest version of Aster can be found at \citet{Github}. In order to install and run the code, users are encouraged to clone the repository using the source control software \texttt{git}; subsequently, one can install the library at the system level using a combination of \texttt{CMake} and \texttt{Make}. Additionally, Aster requires some external libraries to render simulations correctly, thanks to \texttt{OpenGL}.
compile and run kernels on dedicated hardware with \texttt{OpenCL}, and efficiently parallelize many parts of its code with the help of \texttt{TBB}.

\section{Gravity Solvers}
        
Aster provides two main gravitational force evaluators: a simple, but heavily parallelized, gravity solver using a direct $N^2$ algorithm useful for few-body simulation where long-term, near-machine-precision accuracy is essential, and a more advanced spatial tree solver with logarithmic time complexity based on \citet{Barnes1986} and parallelized using Morton encoding. Both solvers have been implemented on CPU and on dedicated hardware.  
		
\subsection{Classical Solver}
The first, more direct, solver is available under the \texttt{PARALLEL} simulation type; it iterates over all the bodies inside the system and subsequently computes their acceleration (if \texttt{force_type} \allowbreak is set 
        
        to \texttt{newtonian})  using
		\begin{dmath*}
			\vec{F}_i =  - \sum^{N}_{j\neq i} G\frac{m_im_j}{|r_{ij}|^3}\vec{r}_{ij}
		\end{dmath*}
Where the inverse root of the distance can be softened with the help of a user-defined constant, $\epsilon$  which serves the purpose of preventing unrealistic accelerations when r approaches zero. 

This approach is particularly computationally demanding since it requires the processor to perform numerous read operations over non-contiguous regions of memory, yielding an algorithm prone to cache misses, which, combined with the inherent $\mathcal{O} (N^2)$ time complexity, produces an accurate but inefficient gravity solver. However, many parts of its code have been heavily parallelized

and optimized using SIMD operations and the \texttt{TBB} parallelization library, allowing this solver to be employed in near-machine-precision few-body problems without lacking in speed or precision compared to other approaches.

\subsection{Spatial Tree Solver}
In order to improve on the previous method, we can leverage a simple approximation during the computation of the accelerations; in particular, one can arrange all the particles into a hierarchical, tree-like data structure (referred to as a quadtree in two spatial dimensions or an octree in three). Each node in the tree allows distant, clustered groups of bodies to be approximated as a single effective particle carrying the aggregate properties of the group, i.e.,  the total mass of the cluster ($M_{node}$) and its barycenter ($\vec{\textbf{r}}_{gc}$):
		\begin{dmath*}
			M_{node} = \sum_i m_i
		\end{dmath*}
		\begin{dmath*}
			\vec{\textbf{r}}_{gc} = \frac{1}{M_{node}} \sum_i m_i \vec{\textbf{r}}_i
		\end{dmath*}
		where $m_i$ and $\vec{\textbf{r}}_i$ are, respectively, the masses and positions of the particles contained in the given node. 
		
		The original implementation of the algorithm in \citet{Barnes1986} generates a spatial tree by recursively traversing the structure for each body until every particle is assigned a unique leaf or a predetermined recursion limit is hit. Subsequently, the accelerations are computed for each body by progressively descending the spatial tree, treating a node as a pseudo-particle if it satisfies the condition
		\begin{dmath}\label{eq:theta_cond}
			S_i < \theta \cdot |\vec{N_i} - \vec{B_i}|
		\end{dmath}
Where, $N_i$ and $B_i$ represent the position vectors of the given node and body, respectively; $\theta$ denotes the opening angle of the algorithm; and $S_i$ indicates the size of the node.

This approach has several inherently sequential steps, a factor that limits its performance and scalability on dedicated hardware. It follows that the original implementation would greatly benefit from a process to concurrently compute leaves and internal nodes from the mere positions of the examined particles. Such a type of process can be developed by following a parallelized approach based on assigning each body with a spatial identifier \citep{Karras2012, 10.1145/3105762.3105782}, rather than a recursive one, to compute the edges and nodes in the spatial tree we are seeking to assemble. More precisely, in the Aster library, this algorithm was accomplished by employing Morton encoding \citep{Morton1966}, not only linearizing the otherwise recursive algorithm but also preserving spatial locality as described in Algorithm 1. We have introduced the $\delta(i, j)$ function that computes the longest common prefix between the $i$th and the $j$th Morton codes in the sorted list, along with the range\_size and range\_end functions, which respectively compute the size of the range covered by the selected node and the index of the last internal node (or pseudo-particle node) in the selected range. Moreover, we assume that every array associated with the tree nodes (e.g., \texttt{node\_pos}, \texttt{node\_mass}, \texttt{left}) is partitioned as follows: 
\[
\text{node}[i] =
\begin{cases}
\text{pseudo-particle} & \text{if } x < N-1 \\
\text{root node}       & \text{if } x = N \\
\text{internal node}   & \text{otherwise}
\end{cases}
\]

\begin{algorithm}[h]
\caption{Parallel LBVH Tree Construction from Morton Codes}
\label{alg:lbvh_build}
\begin{algorithmic}[1]
\FOR{ $i \in [0, N-2]$ in parallel}

    \STATE $d_L \gets \delta(i, i-1)$
    \STATE $d_R \gets \delta(i, i+1)$

    \STATE $d \gets \mathrm{sign}(d_R - d_L)$
    \STATE $d_{min} \gets \min(d_L, d_R)$

    \textcolor{gray}{// Computes the range size and start}
    \STATE $l_{max} \gets $ range\textunderscore size(i, d)
    \STATE $j \gets i + d$ * range\textunderscore end(i, d) 
    \STATE $d_{node} \gets \delta(i,j)$

    \textcolor{gray}{// Binary search to find split point}
    \STATE $s \gets 0$
    \FOR{$t \gets \lceil(l+1)/2\rceil$ down to $1$}
        \IF{$\delta(i, i + (s+t)d) > d_{node}$}
            \STATE $s \gets s + t$
        \ENDIF
    \ENDFOR

    \STATE $\gamma \gets i + s d$

    \STATE internal $\gets N + i$

    \textcolor{gray}{// Assigns the indices to the correct nodes}

    \STATE left[internal] $\gets \gamma $ if $\min(i,j) = \gamma$ else $N + \gamma$

    \STATE right[internal] $\gets \gamma + 1$ 
    \IF {$\max(i,j) \neq \gamma + 1$}
        \STATE right[internal] $\gets $ right[internal] $+ N$
    \ENDIF
    \STATE idx $\gets$ morton[i].index
    \STATE node\_pos[i] $\gets$ positions[idx]
    \STATE node\_mass[i] $\gets$ masses[idx]

\ENDFOR

\end{algorithmic}
\end{algorithm}

\begin{algorithm}[h]
\caption{Barnes--Hut Force Evaluation (3-D, GPU Kernel)}
\label{alg:barnes_hut_force}
\begin{algorithmic}[1]

\FOR{$i \in [0, N)$ in parallel}
    \STATE $my\_pos \gets \mathbf{p}_{i}$
    \STATE $my\_mass \gets m^{(body)}_{i}$
    \STATE $acc \gets \mathbf{0}$
    \STATE $stack.\textsc{push}(\langle root\_node,\; tree\_size \rangle)$
    \WHILE{$\neg\, stack.\textsc{empty}()$}
        \STATE $\langle node,\; size \rangle \gets stack.\textsc{pop}()$
        \STATE \textbf{if} $node \notin [0,\; max\_nodes)$ \textbf{continue}
        \STATE $node\_pos \gets \mathbf{c}_{node}$
        \STATE $d\_sq \gets \|my\_pos - node\_pos\|^2 + \epsilon^2$
        \STATE $is\_leaf \gets (lefts[node] = -1)$
        
        \IF{$size^2 / d\_sq < \theta$ \OR $is\_leaf$}
            \IF{$d\_sq > 0$}
                \STATE $acc \gets acc + \textsc{force}( my\_mass, my\_pos,\, node)$
            \ENDIF
        \ELSE
            \IF{$rights[node] \neq -1$}
                \STATE$stack.\textsc{push}(\langle rights[node],\; size/2 \rangle)$
            \ENDIF
            
            \IF{$lefts[node] \neq -1$}{
                \STATE$stack.\textsc{push}(\langle lefts[node],\; size/2 \rangle)$}
            \ENDIF
        \ENDIF
    \ENDWHILE
    \STATE $acc\_out[i] \gets acc$
\ENDFOR

\end{algorithmic}
\end{algorithm}
After the tree is assembled and each node is assigned the cumulative mass and center of mass from its children, all the interactions are computed using a descent algorithm outlined by Algorithm 2; here, we assign each body to a core on the GPU, which will then repeatedly use Eq. \eqref{eq:theta_cond} to compare distance and size of the cluster to the opening parameter $\theta$ to either descend further or approximate the entire branch as a single pseudo-particle.   
In essence, this particular implementation can be summarized in the following key steps:
		
		\hspace{1em}\small \textbf{1)} Each body is assigned a Morton code based 
		
		\hspace{2.5em}on its coordinates.
		
		\hspace{1em}\small \textbf{2)} The list of Morton codes is sorted.
		
		\hspace{1em}\small \textbf{3)} A binary tree is built from the sorted list (Algorithm 1).
		
		\hspace{1em}\small \textbf{4)} The total mass and barycenter are computed
		
		\hspace{2.5em}for each node in the tree.
		
		\hspace{1em}\small \textbf{5)} Each body’s acceleration is updated based on the tree (Algorithm 2).

Furthermore, in the first step a 32-bit code is assigned to each body by rescaling, normalizing, and interleaving its spatial coordinates. This step is essential to facilitating the parallelization of the construction of the binary radix tree since we can assign the longest shared common prefix among a group of morton codes to each internal node. Subsequently, the list of codes is sorted, and a binary radix tree is built from the sorted list. 
	
\section{Built-in Updaters}

\begin{figure*}[t]
  \centering
    \includegraphics[width=\linewidth]{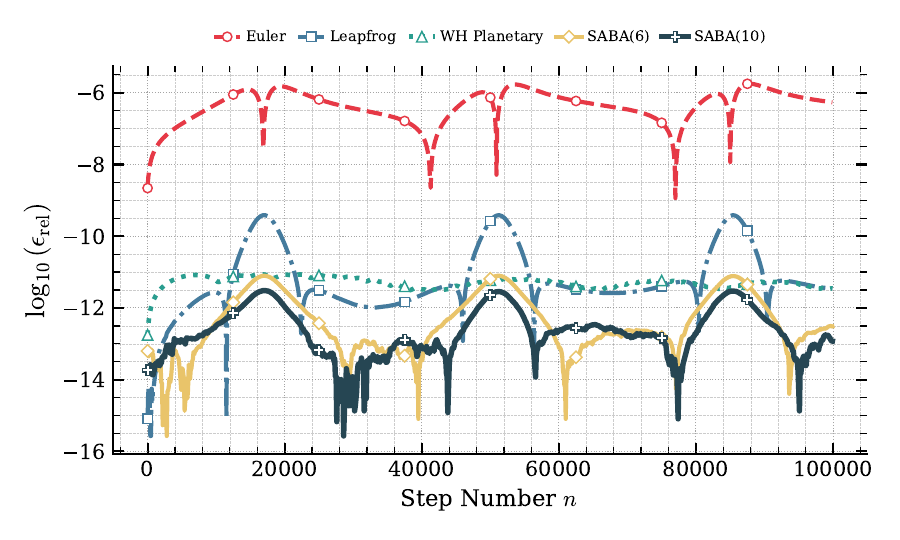}
    \caption{The relative Hamiltonian error of various numerical integrators implemented in the Aster library as it evolves through time. Integrators resulting from the splitting of the Hamiltonian into its kinetic and potential parts (i.e., Leapfrog (blue dash-dotted line), SABA5 (yellow continuous line), and SABA10 (purple continuous line)) follow a similar pattern, oscillating over several orders of magnitude; meanwhile, WH (dotted green line) integration results in a steadier approximation over time. It should be mentioned that the system being integrated consists of two bodies with negligible masses orbiting a larger central mass, a regime where WH integration is particularly suitable.}
  \label{fig:integrator_times}
\end{figure*}
        
The N-body problem is notoriously difficult to evolve numerically due to the inherent cumulative discrepancies associated with rounding errors on the hardware side and long-term energy drift due to the use of low-order integrators or a poor choice of the evolution step size. 

Most of the numerical integrators implemented in the Aster library rely heavily on the process of Hamiltonian splitting, a method for constructing integration schemes by separating the Hamiltonian of a given system $\mathcal{H}$ into two or more terms
		\begin{dmath}
			\mathcal{H} = \mathcal{H}_1 + \mathcal{H}_2
		\end{dmath}
		which implies 
		\begin{dmath}
			\frac{dz}{dt} = L_{H_1}z + L_{H_2}z
		\end{dmath}
		
where $L_H$ denotes the Liouville operator $L_H f = \{f, H\}$ and z is the state vector (p, q). If the two terms, $H_1$ and $H_2$, were to commute (such that $ \{\mathcal{H}_1, \mathcal{H}_2\} = 0 $ using Poisson's bracket notation), then one would be able to exactly evolve the system over a single timestep $\Delta t$ using 
\begin{dmath}
	z(t + \Delta t) = e^{\Delta t L_H} z(t)
\end{dmath}
almost every Hamiltonian splitting associated with a non-trivial N-Body problem does not satisfy said condition and we are therefore forced to use an approximation, which in many cases is based on the Baker-Campbell-Hausdorff (BCH) formula \allowbreak \citep{blanes2008splittingcompositionmethodsnumerical}, yielding
		\begin{dmath}\label{eq:firstorder_symplectic}
			z(t + \Delta t) = e^{\Delta t L_{H_1}} e^{\Delta t L_{H_2}} z(t) + \mathcal{O}(\Delta t^2)
		\end{dmath}
		
\subsection{Symplectic Euler}
One of the simplest and most general approaches to splitting a system’s Hamiltonian would be to partition said system into its canonical coordinate-dependent part and a generalized momentum-dependent part, as such:
	\begin{dmath}
		\mathcal{H} = K(p) + U(q)
	\end{dmath}
Following Eq.\eqref{eq:firstorder_symplectic}, we can approximate to first order the solution to this particular splitting as 
		\begin{center}
			$ z(t + \Delta t) = e^{(L_K + L_U) \Delta t} z(t) \approx e^{L_K \Delta t} e^{L_U \Delta t}z(t) $
		\end{center}
if we then consider the terms $e^{\Delta t L_U}$ and $e^{\Delta t L_K}$ as two separate systems that can be solved sequentially; we are left with an integration scheme better known as symplectic Euler, implemented in the Aster library under the \texttt{EULER} integrator type. 
		
\subsection{Leapfrog Integration}
Furthermore, if we were to apply Strang splitting, a technique that provides a second-order approximation to Eq. \eqref{eq:firstorder_symplectic} by halving the time step, to the symplectic Euler, we arrive at an integration scheme of the form
	\begin{dmath}
		z(t + \Delta t) = e^{\frac{\Delta t }{2} L_K} e^{\Delta t L_U} e^{\frac{\Delta t }{2} L_K}z(t)
	\end{dmath}
which leads to a symplectic, reversible, and stable integration method with a global error $\propto \mathcal{O}(\Delta t^2) $. This integrator is also known as Leapfrog or velocity Verlet and can be accessed in the library with the \texttt{LEAPFROG} integrator type.
		
\subsection{Wisdom-Holman Mapping}
In the Aster library a symplectic Wisdom-Holman mapping is also implemented under the \texttt{WH_PLANETARY} integrator. This type of integrator assumes that the system's potential is dominated by a central mass such that
	\begin{dmath}
		m_0 \gg  \sum_{i = 1}^{N-1} m_i 
	\end{dmath}
where $m_0$ is the mass of the central object (star) and $m_{i>0}$ is the mass of the $i$th external object (planet).
Therefore, one could separate the system's Hamiltonian into a Planet-Planet interactions part and a Planet-Star \citep{Rein2012} part as follows: 
		\begin{dmath}
			\mathcal{H} = \frac{1}{2}\sum_{i=0}^{N-1} \frac{p_i^2}{m_i} - G\sum_{i=0}^{N-1} \sum_{j = i+1}^{N-1} \frac{m_i m_j}{|\vec{r}_{ij}|}
		\end{dmath}
		we can add and subtract to eq. 9 the term
		\begin{dmath*}
			G\sum_{i=1}^{N-1} \frac{m_i m_0}{|\vec{r}_i'|}
		\end{dmath*}
		to obtain
		\begin{dmath}
			\mathcal{H}=\frac{p_0^{\prime 2}}{2 m_0^{\prime}}+\underbrace{\sum_{i=1}^{N-1} \frac{p_i^{\prime 2}}{2 m_i^{\prime}}-\sum_{i=1}^{N-1} \frac{G m_i' M_i}{|   \vec{r}_i'|}}_{\mathcal{H}_{\text{PS}}} +\underbrace{\sum_{i=1}^{N-1} \frac{G m_i^{\prime} M_i}{|\vec{r}_i'|}-\sum_{i=0}^{N-1} \sum_{j=i+1}^{N-1} \frac{G m_i m_j}{|\vec{r}_{ji}|}}_{\mathcal{H}_{\text{PP}}}
		\end{dmath}
\begin{figure}[t]
  \centering\hspace{-3em}
    \includegraphics[width=\columnwidth]{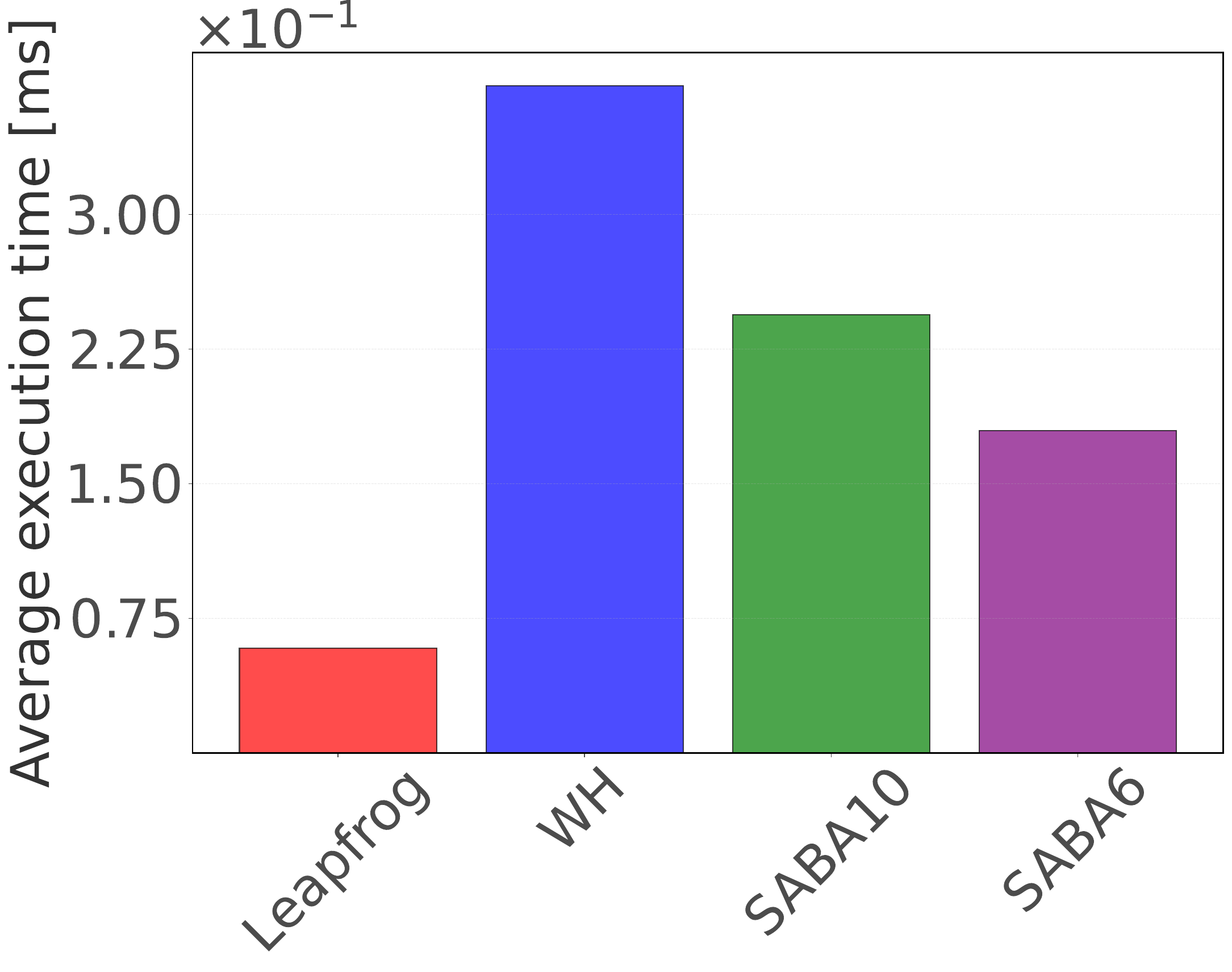}
    \caption{Aster's Barnes-Hut algorithm execution time over the amount of bodies on both CPU and GPU compared to the popular library "Rebound".}
  \label{fig:integrator_times}
\end{figure}

		Here, $\mathcal{H}_{PS}$ refers to the planet-star interactions, which can be solved using Kepler's equations, and $\mathcal{H}_{PP}$ represents planet-planet interactions. Note that we have also introduced the notion of Jacobi coordinates, where
		\begin{dmath*}
			\vec{r}_i' = \vec{r}_i - \vec{R}_{i-1}
		\end{dmath*}
		\begin{dmath*}
			\vec{R_i} = \frac{1}{M_i} \sum_{j=0}^i m_j\vec{r_j}
		\end{dmath*}
		\begin{dmath*}
			M_i = \sum_{j=0}^i m_j
		\end{dmath*}
		\begin{dmath*}
			m_i'=m_i \frac{M_{i-1}}{M_i}
		\end{dmath*}
		
		From eq. 10 we can deduce an integration scheme that best fits our scenario and is therefore more accurate over long time scales:
		\begin{dmath*}
			z(t + \Delta t) = e^{\frac{\Delta t}{2} L_{PS}}e^{\Delta t L_{PP}}e^{\frac{\Delta t}{2} L_{PS}}z(t)
		\end{dmath*}
		algorithmically, it means that for each timestep $\Delta t$ we would have to
	
	\hspace{1em}\small \textbf{1)} Solve Kepler's equations for each body with index $\neq$ 0 
	
	\hspace{2.5em} and evolve for $\frac{\Delta t}{2}$
	
	\hspace{1em}\small \textbf{2)} Compute Planet-Planet interactions and evolve for a full 
	
	\hspace{2.5em}timestep
	
	\hspace{1em}\small \textbf{3)} Solve Kepler's equations for each body with index $\neq$ 0 
	
	\hspace{2.5em} and evolve for $\frac{\Delta t}{2}$
	
	Note that the last step can be merged with the first when the system is evolved for more than one time step $\Delta t$ without any substantial loss in precision. 
	
		\subsection{SABA Integrators}
The fourth, and last, type of numerical integrator implemented in the Aster library is SABA integrators. Similarly to Leapfrog and the symplectic Euler, it is based on the splitting of the Hamiltonian into a momentum-dependent part (A) and a canonical position-dependent part (B) such that 
\begin{dmath*}
	\mathcal{H} = A(p) + B(q)
\end{dmath*}
Then the resulting integration scheme, to first order, is 
\begin{dmath}\label{eqn:SABA_S_n}
	z(t + \Delta t) = e^{\Delta t L_A}e^{\Delta t L_B} z(t) + \mathcal{O}(\Delta t^2)
\end{dmath}
A higher-order approximation can be obtained by applying the BCH formula, leading to the general scheme
\begin{dmath*}
	S_n(\tau) = e^{c_1 \tau L_A}e^{d_1 \tau L_B} \ldots e^{c_n \tau L_A}e^{d_n \tau L_B}
\end{dmath*}
where n is the order of the integration scheme, S is an operator such that 
\begin{dmath*}
	z(t + \Delta t) = S_n(\Delta t) z(t) + \mathcal{O}(t^{n+1})
\end{dmath*}
and we have introduced the coefficients $c_n$ and $d_n$ which are chosen as described in \citet{Laskar2001}

Furthermore, we will only consider symplectic integrators that satisfy $S_n(t)^{-1} = S_n(-t)$, thus the general formula for this type of integrators $SABA_{2n}$ is going to be  
\begin{dmath*}
	e^{c_1 \tau L_A}e^{d_1 \tau L_B} \ldots e^{d_n \tau L_B}e^{c_{n+1} \tau L_A}e^{d_n \tau L_B} \ldots e^{d_1 \tau L_B}e^{c_1 \tau L_A}
\end{dmath*}
meanwhile for $SABA_{2n+1}$
\begin{dmath*}
	e^{c_1 \tau L_A}e^{d_1 \tau L_B} \ldots e^{c_{n+1} \tau L_A}e^{d_{n+1} \tau L_B}e^{c_{n+1} \tau L_A} \ldots e^{d_1 \tau L_B}e^{c_1 \tau L_A}
\end{dmath*}
Note that the combination of symmetry and a higher order of integration allows the SABA family to be reversible in time and particularly effective for long-term integrations.

\section{Comparison With Popular N-body Codes}

To accurately evaluate Aster's performance and accuracy relative to other popular N-body codes, we conduct several tests probing Aster's capabilities at high-energy regimes, and we compare these results against Rebound \citep{Rein2012} and Gadget4 \citep{2021MNRAS.506.2871S}. All benchmarks are performed on an NVIDIA GeForce MX130 for GPU-related tests and an Intel i5-8250U (with a clock speed of 3.4 GHz) for CPU tests. 

\subsection{Execution Time And Accuracy Comparison}
We probe the execution time of Aster, Rebound, and Gadget4 by integrating a self-gravitating disk comprised of a central mass $1M_{\odot}$ and an accretion disk of variable particle count ranging from $N = 10^3$ to $N=10^5$. The disk has an inner radius of $r_{inner} = 2\times10^9m$ and extends radially to $r_{outer} = 4\times10^{11}m$, each particle is initialized with Keplerian velocities. Each library evolves the system using a tree-based Barnes-Hut algorithm with an opening angle $\theta = 0.45$. In this comparison each library makes use of a Leapfrog integrator (or an equivalent second-order integration scheme). 

\begin{figure}[!htbp]
  \centering\includegraphics[width=1 \columnwidth]{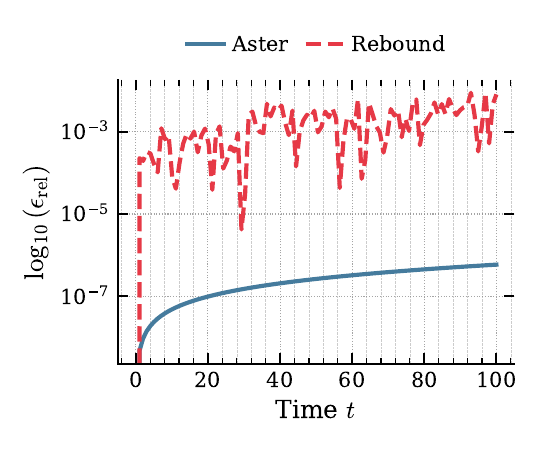}
    \caption{Relative energy error over time of the integration of a self-gravitating disc of $1 \times 10^5$ bodies using a SABA3 integrator on both Aster (in a blue continuous line) and Rebound (red dashed line) evolved over 100 timesteps}
  \label{fig:energy_disc}
\end{figure}

\begin{figure}[t]
  \centering
  \includegraphics[width=1 \columnwidth]{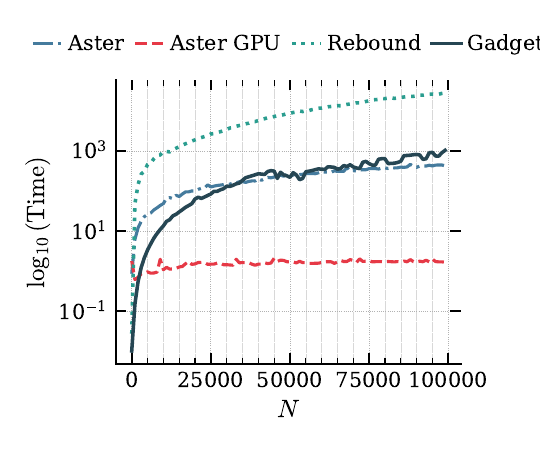}
    \caption{Aster's Barnes-Hut algorithm execution time over the amount of bodies on both CPU (blue dash-dotted) and GPU (red dashed) compared to the popular libraries Rebound (green dotted) and Gadget4 (purple continuous) using comparable parameters.}
  \label{fig:BH_times}
\end{figure}

Figure \ref{fig:BH_times} shows the time of execution as a function of the number of bodies for all three codes. As expected from the algorithmic complexity nature of the tree-based solvers, each library exhibits $\mathcal{O}(Nlog(N))$ scaling behavior; more precisely, we can infer that Aster's approach becomes particularly advantageous when $N\approx 6\cdot 10^4$. This performance gain is due to the library's Morton encoding algorithm that preserves both spatial and data locality, lowering the chance of cache misses, and the linearized, recursion-free routine that results from it, which also contributes to lowering the overhead, significantly improving performance.

Gadget4 exhibits similar execution times to those of Aster across a wide range of particle counts thanks to its greatly optimized TreePM implementation. Nevertheless, Gadget4 is designed for cosmological-scale simulations with periodic boundary conditions, an area where it excels in performance.

To accurately assess the numerical precision of both Aster and Rebound integration methods, we evolve a self-gravitating disk comprised of $10^4$ bodies and a central mass of $10^2M_\odot$. The system is evolved using of a BH solver with a lower opening angle of $\theta = 0.4$ to reduce force truncation errors and integrated using Aster's SABA3 integrator against Rebound's equivalent third-order scheme. 
Figure \ref{fig:energy_disc} shows the relative Hamiltonian error (computed using  $\frac{\Delta E}{E_0} = \frac{|E(t) -E_0|}{E_0}$) over time produced by Aster and Rebound while integrating the system. Both libraries maintain bounded energy errors during the integration period following a pseudo-logarithmic curve due to the spatial tree approximation being employed. with Aster accumulating error logarithmically, maintaining a numerical accuracy comparable to other popular N-body codes for large-scale simulations.

\begin{figure}[H]
  \centering
    \includegraphics[width=1 \columnwidth]{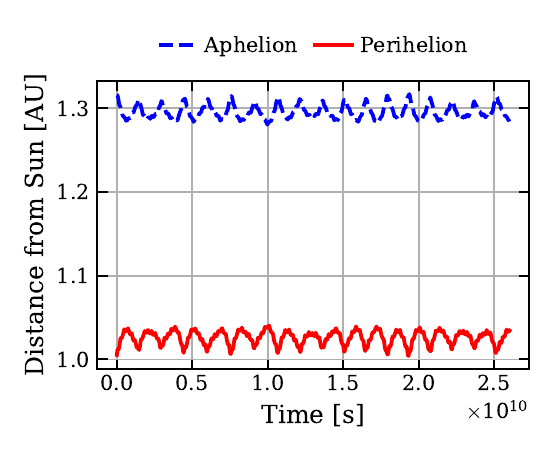}
    \caption{Aphelion (dotted blue line) and perihelion (in red) of Earth plotted over time. This graph shows how Aster is capable of integrating systems over long periods of time, maintaining numerical accuracy. The oscillations and irregularities in the plot are caused by the interaction with other massive bodies in the system (namely Jupiter and Saturn) and the numerical error accumulation.}
  \label{fig:earth_ecc}
\end{figure}

\section{N-body Force Calculation}
The Aster library offers built-in support for post-Newtonian approximations (PNs), including 1PN, 2PN, 2.5PN, and arbitrary user-defined force calculation methods to be employed both on CPU and dedicated hardware. In addition, these PNs can be turned on and easily combined with one another to precisely select different post-Newtonian effects. 

\subsection{Post-Newtonian Approximations}

During the numerical computation of the acceleration acting on a pair of massive particles (which we will label as 1 and 2 for simplicity), such higher-order approximations can be written as a series expansion in powers of v/c (or 1/c), the parameters of which can be found as described in \citet{Blanchet2006}

		\begin{dmath*}
			a=\underbrace{a_0}_{\text {Newt. }}+\underbrace{c^{-2} a_2}_{1 \mathcal{P} \mathcal{N}}+\underbrace{c^{-4} a_4}_{2 \mathcal{P} \mathcal{N}}+\underbrace{c^{-5} a_5}_{2.5 \mathcal{P} \mathcal{N}}+\mathcal{O}\left(c^{-6}\right)
		\end{dmath*}
		Furthermore, we can denote the mass of the two particles as $m_1$ and $m_2$, their respective velocities as $v_1$ and $v_2$, G the gravitational constant, and n the unit vector from particle 2 to particle 1:

		\begin{dmath*}
			a_2 = \frac{G m_2}{r^2}\left(u \left[-v_1^2-2 v_2^2+4 v_1 v_2+\frac{3}{2}\left(n v_2\right)^2+ 5\frac{G m_1}{r}+4\frac{G m_2}{r}\right]+\left(v_1-v_2\right)\left[4 n v_1-3 n v_2\right]\right)
		\end{dmath*}

        \begin{figure}[!htbp]
  \centering
    \includegraphics[width=1 \columnwidth]{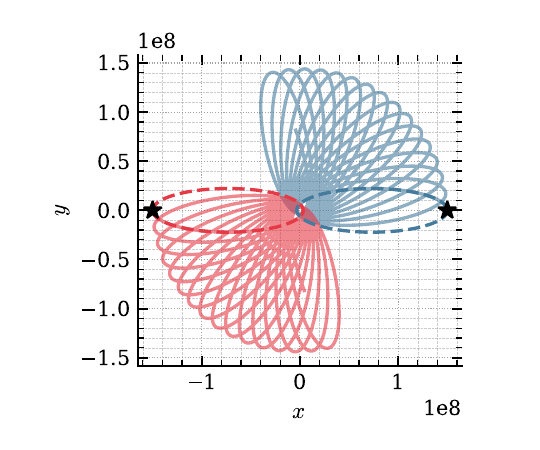}
    \caption{The trajectories of two compact bodies with a mass of $5 M_{\odot}$ and traveling at a speed of $0.0017c$, both considering post-Newtonian terms (faint trajectory lines) and using Newton's law of gravitation alone (dashed trajectories). The starting position of the bodies is denoted by a black star marker. The continuous trajectory portrays precession of the perihelion is noticeable as expected}
  \label{fig:pns_precession}
\end{figure}

		\begin{dmath*}
			a_4=\frac{G m_2}{r^2}\left(u \left[-2 v_2^4+4 v_2^2\left(v_1 v_2\right)-2\left(v_1 v_2\right)^2
			+\frac{3}{2} v_1^2\left(n v_2\right)^2+\frac{9}{2} v_2^2\left(n v_2\right)^2-6\left(v_1 v_2\right)\left(n v_2\right)^2 
			-\frac{15}{8}\left(n v_2\right)^4+\frac{G m_1}{r}(-\frac{15}{4} v_1^2+\frac{5}{4} v_2^2-\frac{5}{2} v_1 v_2
			+\frac{39}{2}\left(n v_1\right)^2-39\left(n v_1\right)\left(n v_2\right)  \allowbreak +\frac{17}{2}\left(n v_2\right)^2) 
			+\frac{G m_2}{r}\left(4 v_2^2-8 v_1 v_2+2\left(n v_1\right)^2
			-4\left(n v_1\right)\left(n v_2\right)-6\left(n v_2\right)^2\right)\right] 
			+\left(v_1-v_2\right)\left[v_1^2\left(n v_2\right)+4 v_2^2\left(n v_1\right)-5 v_2^2\left(n v_2\right)
			-4\left(v_1 v_2\right)\left(n v_1\right)+4\left(v_1 v_2\right)\left(n v_2\right)-6\left(n v_1\right)\left(n v_2\right)^2 
			+\frac{9}{2}\left(n v_2\right)^3+\frac{G m_1}{r}\left(-\frac{63}{4} n v_1+\frac{55}{4} n v_2\right) 
			+\frac{G m_2}{r}\left(-2 n v_1-2 n v_2\right)\right]\right) 
			+\frac{G^3 m_2}{r^4} u\left[-\frac{57}{4} m_1^2-9 m_2^2-\frac{69}{2} m_1 m_2\right]
		\end{dmath*}
		\begin{dmath*}
			a_5= \frac{4}{5} \frac{G^2 m_1 m_2}{r^3}\left(( v _ { 1 } - v _ { 2 } ) \left[-\left(v_1-v_2\right)^2+2\frac{G m_1}{r}
			-8\frac{G m_2}{r}\right] 
			+a\left(n v_1-n v_2\right)\left[3\left(v_1-v_2\right)^2-6\frac{G m_1}{r}+\frac{52}{3}\frac{G m_2}{r}\right]\right)
		\end{dmath*}

In the Aster library, these post-Newtonian terms have been optimized and implemented both on CPU and on dedicated hardware. Since each PN represents a different effect (e.g., self-gravity, precession of the perihelion, and gravitational radiation, as shown in Fig. \ref{fig:pns_precession}), one may want to study and isolate a single term; for this reason, PNs can be chained together into a single force calculation routine by performing a bitwise OR operation on their identifier (\texttt{NEWTON}, \texttt{PN1}, \texttt{PN2}, \texttt{PN25}) and setting the \texttt{force_law} parameter of a Simulation object to the result of said operation.

\subsection{User-defined Computation Routines} In addition to the built-in force calculation routines, Aster also provides its users with the ability to define arbitrary laws of motion. These custom routines introduce terms that act directly on the acceleration of the single particles and can act as a replacement for the built-in force evaluators or as correction terms. For instance, one could easily implement other forces acting on the system, such as atmospheric drag, electromagnetism, or spring-like forces, to account for a broader variety of physical scenarios. It is important to note that pre-existing routines tend to preserve the system's symplectic structure over time, whereas arbitrary force calculation may not comply with these standards and therefore lead to long-term instabilities, especially when paired with a symplectic integrator. 
	
\section{Example use cases}

In this section we present several example use cases in order to probe the library's efficiency and precision over three different regimes, such as a few-body system integrated over large time scales (e.g., the long-term evolution of the solar system), many-body systems dominated by a central potential, and highly elliptic binaries to probe the accuracy and stability of the different numerical integrators implemented in the Aster library. 

\subsection{Long-term Evolution Of Planetary Systems} The long-term evolution of planetary systems is a commonly used benchmark to evaluate the accuracy of a numerical integrator \citep{Rein_2019, Wang_2021} as physically meaningful results require the conservation of energy and angular momentum over extremely long timescales. As shown in \ref{fig:earth_ecc} the Aster is capable of integrating a system over long periods of time while conserving the system's energy and accurately computing the interactions between different orbiting bodies. In particular, the simulation comprised of eight orbiting masses with velocities, positions, and masses comparable to those of the planets in the solar system, and it was evolved over $800$ years using $\Delta t = 864s$ and a SABA integrator of tenth order.

\subsection{Evolution of Galaxy-like Systems} 
\begin{figure}[h]
  \centering
    \includegraphics[width=1 \columnwidth]{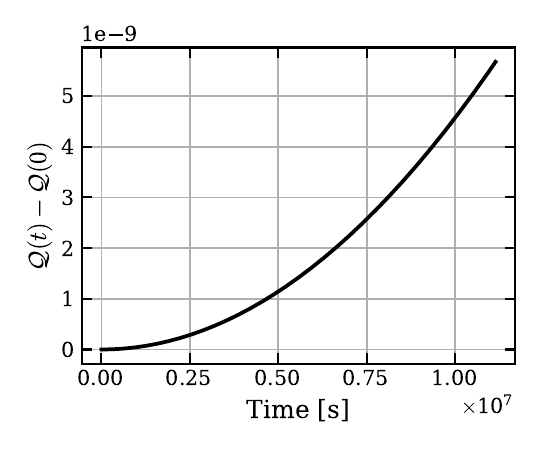}
    \caption{Difference in virial ratio over time for a galaxy-like system as described in paragraph 7.2; the ratio is increasing as a consequence of the numerous interactions between the particles inside the accretion disk}
  \label{fig:ratio}
\end{figure}

\begin{figure}[!htbp]
  \centering
    \includegraphics[width=1 \columnwidth]{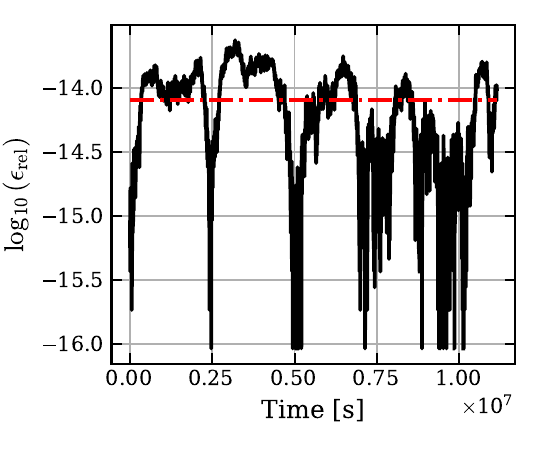}
    \caption{The relative Hamiltonian error of a system dominated by a central mass of $5.03\times10^5 M_{\odot}$ surrounded by an accretion disk comprised of $\approx 10^5$ particles, each with a mass of $1M_{\oplus}$ (black continuous line). The average error over the entire execution is shown as a red dash-dotted line.}
  \label{fig:error_galaxy}
\end{figure}

Simulating the evolution of galaxy-like systems requires computing the gravitational interaction of many particles with high numerical accuracy in order to obtain a physically meaningful result. In this regime, direct summation methods prove to be inefficient; therefore, to accurately probe the library's efficiency and precision, we will make use of a BH gravity solver. To better analyze this system we can make use of a quantity that can be directly derived from the virial theorem, i.e., the virial ratio, defined as 
\begin{dmath*}
    Q(t) = \frac{2\mathcal{K}(t)}{|\mathcal{U}(t)|}
\end{dmath*}
where $\mathcal{K}(t)$ and $\mathcal{U}(t)$ are the kinetic and potential energy of our system at a time t, respectively. More specifically, \citet{refId0} have shown that the stability condition for a gravitationally dominated system 
\begin{dmath*}
    2 \langle\mathcal{K}\rangle + \langle\mathcal{U}\rangle = 0
\end{dmath*}
directly follows from 
\begin{dmath*}
    \langle \frac{dG}{dt} \rangle = 0
\end{dmath*}
hence we can exploit the virial ratio as a valuable resource to determine the tendency of a system to either collapse (when $Q(t) \ll 1$) or expand (when $Q(t) \gg 2$). By employing Aster in the evolution of a gravitationally bound system with an initial virial ratio of $\approx 0.7$. In the particular system we are trying to integrate, we can infer that the cumulative Hamiltonian error remains bounded over the entire integration period (as shown in \ref{fig:error_galaxy}), however, as shown in \ref{fig:ratio}, the virial ratio is constantly increasing over time. This effect can be explained by the interactions taking place inside the accretion disk among the particles, which drives the total kinetic energy up while lowering the potential energy. We can therefore conclude that Aster can be employed in many-body chaotic systems, as it is capable of maintaining a low enough Hamiltonian error to allow for physically meaningful results. 

\subsection{Integration of highly eccentric binaries} 
\begin{figure}[t]
  \centering
    \includegraphics[width=1 \columnwidth]{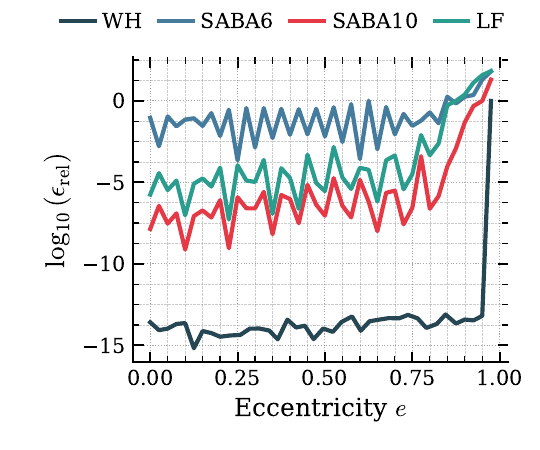}
    \caption{The relative Hamiltonian error of the Wisdom-Holmann (in purple), SABA6 (in blue), SABA10 (in red), and the leapfrog integrator (in azure) plotted over the eccentricity of the system being integrated.  The system is comprised of a central body with a mass of $1 M_{\odot}$ and a smaller orbiting mass of $0.055M_{\oplus}$; the apoapsis  is held fixed at $0.31AU$. The oscillatory behavior displayed by SABA6, SABA10, and WH is intrinsic to integrators constructed from a kinetic/potential splitting and is the result of coupling between the algorithm substep timestep coefficient and the timestep used in the simulation; Wisdom-Holman on the other hand, has a bounded error until an eccentricity of 0.8 due to the different way it splits the Hamiltonian.}
  \label{fig:ecc_prec}
\end{figure}

Highly eccentric binary systems, i.e., two-body systems with an eccentricity exceeding $e = 0.7$, are a particularly challenging class of problems for numerical integrators since they are often accompanied by rapidly varying accelerations (e.g., during close encounters between the two bodies, especially at the periastron). In this example, we employ Aster to numerically integrate various systems with eccentricity varying from perfectly circular to almost parabolic. As shown in  Fig. \ref{fig:ecc_prec} all integrators deriving from the kinetic-potential splitting follow a similar cyclic pattern over the space of possible eccentricities, with each order contributing to a lower numerical error, ultimately degenerating at $e\approx 0.8$, while the $\mathcal{H}_{PP} $- $\mathcal{H}_{PS}$ splitting used in WH integration proves to be more suitable for a wider range of eccentricities, keeping a consistent near-machine precision until $e = 0.96$.

\section{Summary}

In this work we have presented the Aster library, a broad collection of mathematical tools aimed at providing a flexible and reliable environment that can be of great use in studying a vast range of astrophysical problems. Aster's source code is currently being hosted by \citet{Github} under an open-source license, and it can be downloaded using the source control software \texttt{git}. 

Furthermore, we have shown how the library behaves, both performance- and precision-wise, with ad hoc tests based on realistic physical scenarios. The library, in its entirety, is completely written in C++, making use of some external tools, such as \texttt{TBB} for parallelization, \texttt{OpenCL} to run N-body code on dedicated hardware, and \texttt{OpenGL} to provide an intuitive, performant, and customizable user interface.

Additionally, we have presented the two main gravity solvers implemented in the library: a slower, yet precise, few-body solver based on a $\mathcal{O}(N^2)$ approach, and a significantly faster, approximate solver relying on a combination of the classical Barnes-Hut algorithm and Morton encoding, which has proven to be particularly efficient both on CPU and GPU while maintaining a relatively high precision. Lastly, we presented the issues non-symplectic integrators have and how said problems are overcome thanks to the extensive  use of high-order symplectic numerical integration routines in the Aster library and, more specifically, how these algorithms excel in precision when faced with realistic astrophysical scenarios.

Given the aforementioned characteristics, we expect Aster to be employed in studies where numerical accuracy and flexibility are indispensable, both in an astrophysical context and in broader, more theoretical studies. 

\bibliographystyle{IEEEtranN}
\bibliography{main}

@article{Blanchet2006,
  author  = {Blanchet, Luc},
  title   = {Gravitational Radiation from Post-Newtonian Sources and Inspiralling Compact Binaries},
  journal = {Living Reviews in Relativity},
  year    = {2006},
  volume  = {9},
  number  = {1},
  pages   = {4},
  doi     = {10.12942/lrr-2006-4},
  url     = {https://link.springer.com/article/10.12942/lrr-2006-4},
  issn    = {1433-8351}
}

@article{Morton1966,
  author  = {Morton, G. M.},
  title   = {A computer oriented geodetic data base and a new technique in file sequencing},
  journal = {IBM Ltd.},
  year    = {1966},
  note    = {Ottawa, Canada}
}

@misc{blanes2008splittingcompositionmethodsnumerical,
      title={Splitting and composition methods in the numerical integration of differential equations}, 
      author={Sergio Blanes and Fernando Casas and Ander Murua},
      year={2008},
      eprint={0812.0377},
      archivePrefix={arXiv},
      primaryClass={math.NA},
      url={https://arxiv.org/abs/0812.0377}, 
}

@misc{GitHub,
    author = {{GitHub}},
    title = {{Aster}},
    howpublished = {\url{https://github.com/F-Benevelli/Aster}},
    note = {Software repository}
}

@article{Barnes1986,
	author = {Josh Barnes and Piet Hut},
	title = {A hierarchical O(N log N) force-calculation algorithm},
	journal = {Nature},
	year = {1986},
	month = {12},
	volume = {324},
	pages = {446--449},
	url = {https://www.nature.com/articles/324446a0},
}

@article{Karras2012,
	author = {Tero Karras},
	title = {Maximizing Parallelism in the Construction of BVHs, Octrees, and k-d Trees},
	journal = {NVIDIA Research},
	year = {2012},
	url = {https://research.nvidia.com/sites/default/files/pubs/2012-06_Maximizing-Parallelism-in/karras2012hpg_paper.pdf},
}

@article{Rein2012,
	author = {H. Rein and S.-F. Liu},
	title = {REBOUND: an open-source multi-purpose N-body code for collisional dynamics},
	journal = {Astronomy \& Astrophysics},
	year = {2012},
	month = {jan},
	volume = {537},
	pages = {A128},
  url = {https://www.aanda.org/articles/aa/abs/2012/01/aa18085-11/aa18085-11.html},
	doi = {10.1051/0004-6361/201118085},
}

@inproceedings{10.1145/3105762.3105782,
author = {Vinkler, Marek and Bittner, Jiri and Havran, Vlastimil},
title = {Extended Morton codes for high performance bounding volume hierarchy construction},
year = {2017},
isbn = {9781450351010},
publisher = {Association for Computing Machinery},
address = {New York, NY, USA},
url = {https://dl.acm.org/doi/abs/10.1145/3105762.3105782},
doi = {10.1145/3105762.3105782},
booktitle = {Proceedings of High Performance Graphics},
articleno = {9},
numpages = {8},
location = {Los Angeles, California},
series = {HPG '17}
}

@article{Laskar2001,
	author = {Jacques Laskar and Philippe Robutel},
	title = {High order symplectic integrators for perturbed Hamiltonian systems},
	journal = {Celestial Mechanics and Dynamical Astronomy},
	year = {2001},
	month = {05},
	url = {https://link.springer.com/article/10.1023/A:1012098603882},
}

@article{Rein_2019,
   title={High-order symplectic integrators for planetary dynamics and their implementation in rebound},
   volume={489},
   ISSN={1365-2966},
   url={https://academic.oup.com/mnras/article/489/4/4632/5565063},
   DOI={10.1093/mnras/stz2503},
   number={4},
   journal={Monthly Notices of the Royal Astronomical Society},
   publisher={Oxford University Press (OUP)},
   author={Rein, Hanno and Tamayo, Daniel and Brown, Garett},
   year={2019},
   month=sep, pages={4632–4640} }

@article{ refId0,
	author = {{Terebizh, Valery Yu.}},
	title = {Evolution of a stellar system in the context of the virial equation},
	DOI= "10.1051/0004-6361/202348933",
	url= "https://www.aanda.org/articles/aa/abs/2024/06/aa48933-23/aa48933-23.html",
	journal = {A\&A},
	year = 2024,
	volume = 686,
	pages = "A35",
}

@article{Wang_2021,
   title={SpaceHub: A high-performance gravity integration toolkit for few-body problems in astrophysics},
   volume={505},
   ISSN={1365-2966},
   url={https://academic.oup.com/mnras/article/505/1/1053/6261213},
   DOI={10.1093/mnras/stab1189},
   number={1},
   journal={Monthly Notices of the Royal Astronomical Society},
   publisher={Oxford University Press (OUP)},
   author={Wang, Yi-Han and Leigh, Nathan W C and Liu, Bin and Perna, Rosalba},
   year={2021},
   month=apr, pages={1053–1070} 
}

@ARTICLE{2021MNRAS.506.2871S,
       author = {{Springel}, Volker and {Pakmor}, R{\"u}diger and {Zier}, Oliver and {Reinecke}, Martin},
        title = "{Simulating cosmic structure formation with the GADGET-4 code}",
      journal = {Monthly Notices of the Royal Astronomical Society},
         year = 2021,
        month = sep,
       volume = {506},
       number = {2},
        pages = {2871-2949},
          doi = {10.1093/mnras/stab1855},
archivePrefix = {arXiv},
       eprint = {2010.03567},
 primaryClass = {astro-ph.IM},
       adsurl = {https://ui.adsabs.harvard.edu/abs/2021MNRAS.506.2871S}
}

\end{document}